\documentclass[sigconf]{acmart}

\usepackage{balance}

\def\BibTeX{{\rm B\kern-.05em{\sc i\kern-.025em b}\kern-.08emT\kern-.1667em\lower.7ex\hbox{E}\kern-.125emX}}
    
\copyrightyear{2020}
\acmYear{2020}
\setcopyright{acmcopyright}
\acmConference[WSDM '20] {The Thirteenth ACM International Conference on Web Search and Data Mining}{February 3--7, 2020}{Houston, TX, USA}
\acmBooktitle{The Thirteenth ACM International Conference on Web Search and Data Mining (WSDM'20), February 3--7, 2020, Houston, TX, USA}
\acmPrice{15.00}
\acmDOI{10.1145/3336191.3371850}
\acmISBN{978-1-4503-6822-3/20/02}

\begin{document}

\fancyhead{}

\title{HyperML: A Boosting Metric Learning Approach in Hyperbolic Space for Recommender Systems}


\author{Lucas Vinh Tran}
\affiliation{
  \institution{Nanyang Technological University}
  \institution{Institute for Infocomm Research, A*STAR}
}
\email{trandang001@e.ntu.edu.sg}

\author{Yi Tay}
\authornote{Now at Google Research.}
\affiliation{
  \institution{Nanyang Technological University}
}
\email{ytay017@e.ntu.edu.sg}
 
\author{Shuai Zhang}
\affiliation{
  \institution{University of New South Wales}
}
\email{shuai.zhang@student.unsw.edu.au}

\author{Gao Cong}
\affiliation{
  \institution{Nanyang Technological University}
}
\email{gaocong@ntu.edu.sg}

\author{Xiaoli Li}
\affiliation{
  \institution{Institute for Infocomm Research, A*STAR}
}
\email{xlli@i2r.a-star.edu.sg}

%

%
\begin{abstract}
This paper investigates the notion of learning user and item representations in non-Euclidean space. Specifically, we study the connection between metric learning in hyperbolic space and collaborative filtering by exploring M\"{o}bius gyrovector spaces where the formalism of the spaces could be utilized to generalize the most common Euclidean vector operations. Overall, this work aims to bridge the gap between Euclidean and hyperbolic geometry in recommender systems through metric learning approach. We propose HyperML (Hyperbolic Metric Learning), a conceptually simple but highly effective model for boosting the performance. Via a series of extensive experiments, we show that our proposed HyperML not only outperforms their Euclidean counterparts, but also achieves state-of-the-art performance on multiple benchmark datasets, demonstrating the effectiveness of personalized recommendation in hyperbolic geometry. 
\end{abstract}

%
%

\ccsdesc[500]{Information systems~Recommender systems}
\ccsdesc[500]{Computing methodologies~Neural networks}

%
\keywords{Recommender Systems; Collaborative Filtering; Hyperbolic Neural Networks}

%

%
\maketitle

\vspace{-3ex}
\section{Introduction}

A diverse plethora of machine learning models solves the personalized ranking problem in recommender systems via building matching functions \cite{DBLP:conf/uai/RendleFGS09,DBLP:conf/icdm/Rendle10,mnih2008probabilistic,DBLP:conf/www/HeLZNHC17}. Across the literature, a variety of matching functions have been traditionally adopted, such as inner product \cite{DBLP:conf/uai/RendleFGS09}, metric learning \cite{Tay:2018:LRM:3178876.3186154,DBLP:conf/www/HsiehYCLBE17} and/or neural networks \cite{DBLP:conf/www/HeLZNHC17,He:2018:OPN:3304889.3304969}. Among those approaches, metric learning models (e.g., Collaborative Metric Learning (CML) \cite{DBLP:conf/www/HsiehYCLBE17} and Latent Relational Metric Learning (LRML) \cite{Tay:2018:LRM:3178876.3186154}) are primarily focused on designing distance functions over objects (i.e., between users and items), demonstrating reasonable empirical success for collaborative ranking with implicit feedback. Nevertheless, those matching functions only covered the scope of Euclidean space. 

For the first time, our work explores the notion of learning user-item representations in terms of metric learning in hyperbolic space, in which hyperbolic representation learning has recently demonstrated great potential across a diverse range of applications such as learning entity hierarchies \cite{DBLP:conf/nips/NickelK17} and/or natural language processing \cite{DBLP:conf/wsdm/TayTH18,DBLP:conf/textgraphs/DhingraSNDD18}. Due to the exponentially expansion property of hyperbolic space, we discovered that metric learning with the pull-push mechanism in hyperbolic space could boost the performance significantly: moving a point to a certain distance will require a much smaller force in hyperbolic space than in Euclidean space. To this end, in order to perform metric learning in hyperbolic space, we employ M\"{o}bius gyrovector spaces to generally formalize most common Euclidean operations such as addition, multiplication, exponential map and logarithmic map \cite{DBLP:conf/nips/GaneaBH18,DBLP:series/synthesis/2008Ungar}. 

Moreover, the ultimate goal when embedding a space into another is to preserve distances and more complex relationships. Thus, our work also introduces the definition of distortion to maintain good representations in hyperbolic space both locally and globally, while controlling the performance through the multi-task learning framework. This reinforces the key idea of modeling user-item pairs in hyperbolic space, while maintaining the simplicity and effectiveness of the metric learning paradigm.  

We show that a conceptually simple hyperbolic adaptation in terms of metric learning is capable of not only achieving very competitive results, but also outperforming recent advanced Euclidean metric learning models on multiple personalized ranking benchmarks.
\vspace{-3ex}
\paragraph{Our Contributions}
The key contributions of our work are summarized as follows:

\begin{itemize}
\item We investigate the notion of training recommender systems in hyperbolic space as opposed to Euclidean space by exploring M\"{o}bius gyrovector spaces with the Riemannian geometry of the Poincar\'e model. To the best of our knowledge, this is the first work that explores the use of hyperbolic space for metric learning in the recommender systems domain.
\item We devise a new method HyperML (\textit{Hyperbolic Metric Learning}), a strong competitive metric learning model for one-class collaborative filtering (i.e., personalized ranking). Unlike previous metric learning models, we incorporate a penalty term called \textit{distortion} to control and balance between accuracy and preservation of distances. 
\item We conduct a series of extensive experiments delving into the inner workings of our proposed HyperML on \textbf{ten} public benchmark datasets. Our model demonstrates the effectiveness of hyperbolic geometry, outperforming not only its Euclidean counterparts but also a suite of competitive baselines. Notably, HyperML outperforms the state-of-the-art CML and LRML models, which are also metric learning models in Euclidean space across all benchmarks. We achieve a boosting performance gain over competitors, pulling ahead by up to $32.32\%$ performance in terms of standard ranking metrics.
\end{itemize}

\vspace{-2ex}
\section{Hyperbolic Metric Learning}
\label{sec:hyperbolic_metric_learning}

This section provides the overall background and outlines the formulation of our proposed model. The key motivation behind our proposed model is to embed the two user-item pairs into hyperbolic space, creating the gradients of pulling the distance between the positive user-item pair close and pushing the negative user-item pair away. 

\begin{figure*}[t]
\centering
\begin{minipage}[t]{13.0cm}
\includegraphics[width=8.2cm]{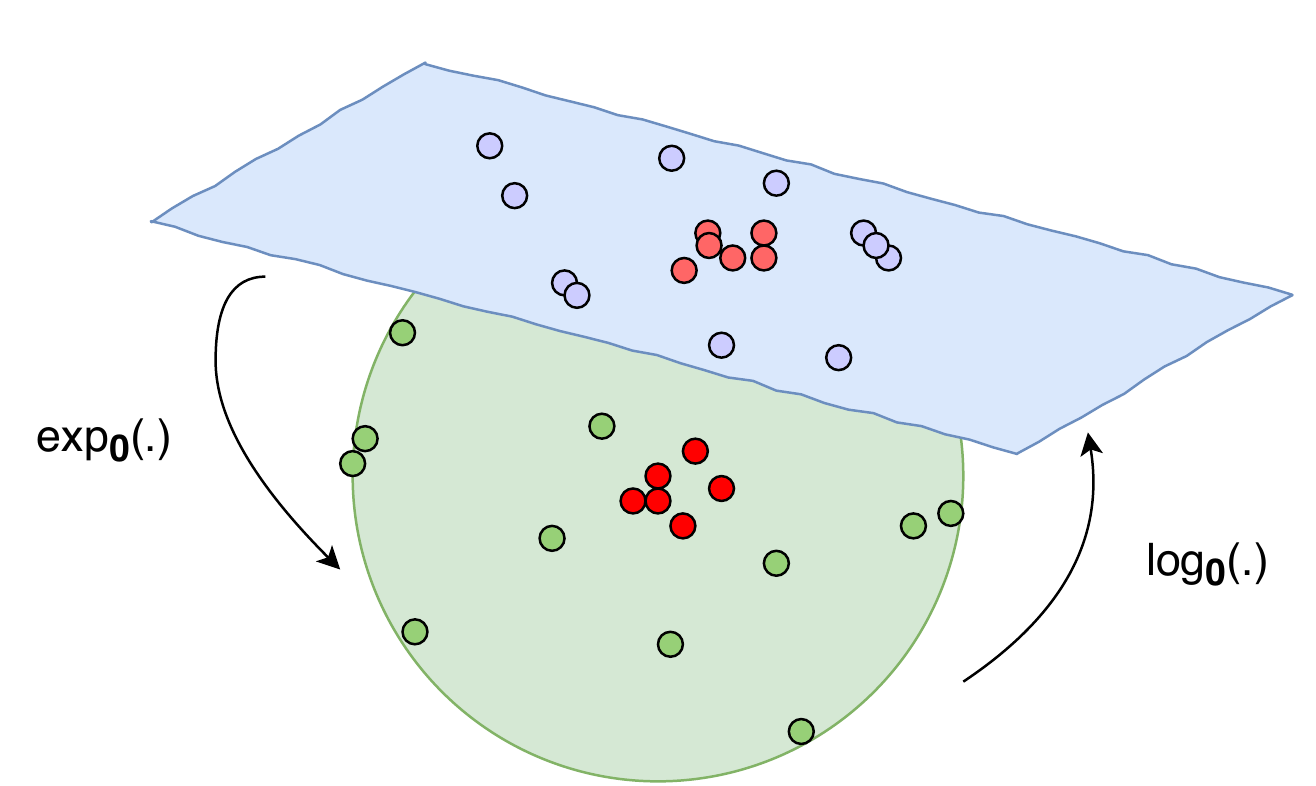}
\includegraphics[width=3.7cm]{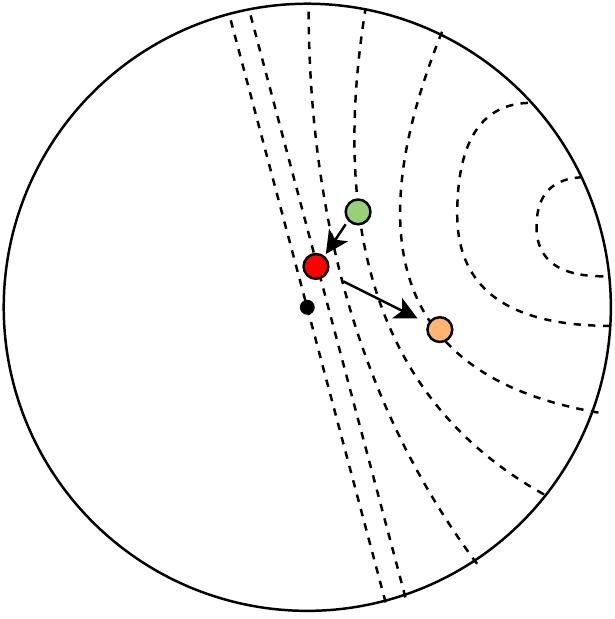}
\end{minipage}
\caption{Illustration of our proposed HyperML. The \textit{left} figure with the greenish ball represents the hyperbolic unit ball and the pale blue parallelogram illustrates its tangent space; red and vermeil circles represent user embeddings; green and purple circles represent item embeddings. The \textit{right} figure illustrates an example of a triplet embedding of user (red circle), positive item (green circle) and negative item (orange circle), in which it demonstrates a small tree of one root and two children which is embedded into hyperbolic space with the exponentially expansion property (\textmd{\textit{Best viewed in color}}).}
\vspace{-3ex}
\label{fig:model}
\end{figure*}

Figure \ref{fig:model} depicts our overall proposed HyperML model. The figures illustrate our two approaches: 1) optimizing the embeddings within the unit ball and 2) transferring the points to the tangent space via the exponential and logarithmic maps for optimization. In the experiments, we also compare the mentioned variants of HyperML where both approaches achieve competitive results compared to Euclidean metric learning models.

\vspace{-2ex}
\subsection{Hyperbolic Geometry \& Poincar\'e Embeddings}
The hyperbolic space $\mathbb{D}$ is uniquely defined as a complete and simply connected Riemannian manifold with constant negative curvature \cite{DBLP:journals/corr/abs-1006-5169}. In fact, there are three types of the Riemannian manifolds of constant curvature, which are Euclidean geometry (constant vanishing sectional curvature), spherical geometry (constant positive sectional curvature) and hyperbolic geometry (constant negative sectional curvature). In this paper, we focus on Euclidean space and hyperbolic space due to the key difference in their space expansion. Indeed, hyperbolic spaces expand faster (exponentially) than Euclidean spaces (polynomially). Specifically, for instance, in the two-dimensional hyperbolic space $\mathbb{D}^2_\epsilon$ of constant curvature $K = -\epsilon^2<0$, $\epsilon>0$ with the hyperbolic radius of $r$, we have:
\begin{equation}
\vspace{-1ex}
L(r) = 2\pi \sinh (\epsilon r), \label{length_of_circle}
\end{equation}
\begin{equation}
A(r) = 2\pi (\cosh (\epsilon r) - 1), \label{area_of_disk}
\end{equation}
in which $L(r)$ is the length of the circle and $A(r)$ is the area of the disk. Hence, both equations illustrate the exponentially expansion of the hyperbolic space $\mathbb{H}^2_\epsilon$ with respect to the radius $r$.

Although hyperbolic space cannot be isometrically embedded into Euclidean space, there exists multiple models of hyperbolic geometry that can be formulated as a subset of Euclidean space and are very insightful to work with, depending on different tasks. In this work, we prefer the Poincar\'e ball model due to its conformality (i.e., angles are preserved between hyperbolic and Euclidean space) and convenient parameterization \cite{DBLP:conf/nips/NickelK17}.

The Poincar\'e ball model is the Riemannian manifold $\mathcal{P}^n = (\mathbb{D}^n, g_p)$, in which $\mathbb{D}^n = \{\textbf{x} \in \mathbb{R}^n: \|\textbf{x}\| < 1\}$ is the \textit{open} $n$-dimensional unit ball that is equipped with the metric:
\begin{equation}
\vspace{-1ex}
g_p (\textbf{x}) =  \Bigg(\frac{2}{1 - \|\textbf{x}\|^2}\Bigg)^2 g_e, 
\end{equation}
where $\textbf{x} \in \mathbb{D}^n$; $\|\cdot\|$ denotes the Euclidean norm; and $g_e$ is the Euclidean metric tensor with components $\textbf{I}_n$ of $\mathbb{R}^n$. 

The induced distance between two points on $\mathbb{D}^n$ is given by:
\begin{equation}
d_\mathbb{D} (\textbf{x}, \textbf{y}) =  \cosh^{-1} \Bigg(1 + 2\frac{\|\textbf{x} - \textbf{y}\|^2}{(1 - \|\textbf{x}\|^2)(1 - \|\textbf{y}\|^2)} \Bigg). \label{dist}
\end{equation}

In fact, if we adopt the hyperbolic distance function as a matching function to model the relationships between users and items, the hyperbolic distance $d_\mathbb{D}(\textbf{u}, \textbf{v})$ between user $u$ and item $v$ could be calculated based on Eqn. (\ref{dist}). 

On a side note, let $\textbf{v}_j$ and $\textbf{v}_k$ represent the items user $i$ liked and did not like with $d_\mathbb{D}(\textbf{u}_i, \textbf{v}_j)$ and $d_\mathbb{D}(\textbf{u}_i, \textbf{v}_k)$ are their distances to the user $i$ on hyperbolic space, respectively. Our goal is to pull $\textbf{v}_j$ close to $\textbf{u}_i$ while pushing $\textbf{v}_k$ away from $\textbf{u}_i$. If we consider the triplet as a tree with two children $\textbf{v}_j$, $\textbf{v}_k$ of parent $\textbf{u}_i$ and place $\textbf{u}_i$ relatively close to the origin, the graph distance of $\textbf{v}_j$ and $\textbf{v}_k$ is obviously calculated as $d(\textbf{v}_j, \textbf{v}_k) = d(\textbf{u}_i, \textbf{v}_j) + d(\textbf{u}_i, \textbf{v}_k)$, or we will obtain the ratio $\frac{d(\textbf{v}_j, \textbf{v}_k)}{d(\textbf{u}_i, \textbf{v}_j) + d(\textbf{u}_i, \textbf{v}_k)} = 1$. If we embed the triplet in Euclidean space, the ratio $\frac{d_\mathbb{E}(\textbf{v}_j, \textbf{v}_k)}{d_\mathbb{E}(\textbf{u}_i, \textbf{v}_j) + d_\mathbb{E}(\textbf{u}_i, \textbf{v}_k)}$ is constant, which seems not to capture the mentioned graph-like structure. However, in hyperbolic space, the ratio $\frac{d_\mathbb{D}(\textbf{v}_j, \textbf{v}_k)}{d_\mathbb{D}(\textbf{u}_i, \textbf{v}_j) + d_\mathbb{D}(\textbf{u}_i, \textbf{v}_k)}$ approaches 1 as the edges are long enough, which makes the distances nearly preserved \cite{DBLP:conf/icml/SalaSGR18}. 

Thus, it is worth mentioning our key idea is that for a given triplet, we aim to embed the root (the user) arbitrarily close to the origin and space the children (positive and negative items) around a sphere centered at the parent. Notably, the distances between points grows exponentially as the norm of the vectors approaches 1.  Geometrically, if we place the root node of a tree at the origin of $\mathbb{D}^n$, the children nodes spread out exponentially with their distances to the root towards the boundary of the ball due to the above mentioned property.

\subsection{Gyrovector spaces}
In this section, we make use of M\"{o}bius gyrovector spaces operations \cite{DBLP:conf/nips/GaneaBH18} to generally design the distance of user-item pairs for further extension.

Specifically, for $c \geq 0$, we denote $\mathbb{D}_c^n = \{x \in \mathbb{R}^n: c\|x\|^2 < 1\}$, which is considered as the open ball of radius $\frac{1}{\sqrt{c}}$. Note that if $c=0$, we get $\mathbb{D}_c^n = \mathbb{R}^n$; and if $c=1$, we retrieve the usual unit ball as $\mathbb{D}_c^n = \mathbb{D}^n$.

Some widely used M\"{o}bius operations of gyrovector spaces are introduced as follows:

\paragraph{M\"{o}bius addition:} The M\"{o}bius addition of $x$ and $y$ in $\mathbb{D}_c^n$ is defined: 
\begin{equation}
x \oplus_c y = \frac{(1 + 2c\langle x,y \rangle + c\|y\|^2)x + (1 - c\|x\|^2)y}{1 + 2c\langle x,y \rangle + c\|x\|^2 \|y\|^2}.
\end{equation}

\paragraph{M\"{o}bius scalar multiplication:} For $c > 0$, the M\"{o}bius scalar multiplication of $x \in \mathbb{D}_c^n \setminus \{\textbf{0}\}$ with $r \in \mathbb{R}$ is defined: 
\begin{equation}
r \otimes_c x = \frac{1}{\sqrt{c}} \tanh(r \tanh^{-1}(\sqrt{c}\|x\|)) \frac{x}{\|x\|},
\end{equation}
and $r \otimes_c \textbf{0} = \textbf{0}$. Note that when $c \rightarrow 0$, we recover the Euclidean addition and scalar multiplication. The M\"{o}bius subtraction can also be obtained as $x \ominus_c y = x \oplus_c (-y)$. 

\paragraph{M\"{o}bius exponential and logarithmic maps:} For any $x \in \mathbb{D}_c^n$, the M\"{o}bius exponential map and logarithmic map, given $v \neq \textbf{0}$ and $y \neq x$, are defined: 
\begin{equation}
\exp_x^c (v) = x \oplus_c \Big(\tanh \Big(\sqrt{c}\frac{\lambda_x^c \|v\|}{2}\Big) \frac{v}{\sqrt{c} \|v\|} \Big),
\end{equation}
\begin{equation}
\log_x^c (y) = \frac{2}{\lambda_x^c \sqrt{c}} \tanh^{-1}(\sqrt{c}\|(-x) \oplus_c y\|) \frac{(-x) \oplus_c y}{\|(-x) \oplus_c y\|},
\end{equation}
where $\lambda_x^c = \frac{2}{1 - c\|x\|^2}$ is the conformal factor of $(\mathbb{D}_c^n, g^c)$ in which $g^c$ is the generalized hyperbolic metric tensor. We also recover the Euclidean exponential map and logarithmic map as $c \rightarrow 0$. Readers can refer to \cite{DBLP:conf/nips/GaneaBH18,DBLP:series/synthesis/2008Ungar} for the detailed introduction to Gyrovector spaces. 

We then obtain the generalized distance in Gyrovector spaces:
\begin{equation}
d_c(x, y) = \frac{2}{\sqrt{c}} \tanh^{-1}(\sqrt{c}\|(-x) \oplus_c y\|).
\end{equation}

When $c \rightarrow 0$, we recover the Euclidean distance since we have $\lim_{c \to 0} d_c(x, y) = 2\|x - y\|$. When $c = 1$, we retrieve the Eqn. (\ref{dist}). In other words, hyperbolic space resembles Euclidean as it gets closer to the origin, which motivates us to design our loss function in the multi-task learning framework.
\subsection{Model Formulation}

Our proposed model takes a user (denoted as $\textbf{u}_i$), a positive (observed) item (denoted as $\textbf{v}_j$) and a negative (unobserved) item (denoted as $\textbf{v}_k$) as an input. Each user and item is represented as a one-hot vector which map onto a dense low-dimensional vector by indexing onto an user/item embedding matrix. We learn these vectors with the generalized distance: 
\begin{equation}
d_c(\textbf{u}, \textbf{v}) = \frac{2}{\sqrt{c}} \tanh^{-1}(\sqrt{c}\|(-\textbf{u}) \oplus_c \textbf{v}\|),
\end{equation}
in which an item $j$ that user $i$ liked (positive) is expected to be closer to the user than the ones he did not like (negative). 

In fact, we would like to learn the user-item joint metric to encode the observed positive feedback. Specifically, the learned metric pulls the positive pairs closer and pushes the other pairs further apart.

Notably, this process will also cluster the users who like the same items together, and the items that are liked by the same users together, due to the triangle inequalities. Similar to \cite{DBLP:conf/www/HsiehYCLBE17}, for a given user, the nearest neighborhood items are: 1) the items liked by this user previously, and 2) the items liked by other users with similar interests to this user. In other words, we are also able to indirectly observe the relationships between user-user pairs and item-item pairs through the pull-push mechanism of metric learning.

\paragraph{Pull-and-Push Optimization.} To formulate such constraint, we define our pull-and-push loss function as:
\begin{equation}
\mathcal{L}_{P} = \sum_{(i, j) \in \mathbb{S}} \sum_{(i, k) \notin \mathbb{S}} [m + d_\mathbb{D}^2(i, j) - d_\mathbb{D}^2(i, k)]_+,
\end{equation}
where $j$ is an item user $i$ liked and $k$ is the one he did not like; $\mathbb{S}$ contains all the observed implicit feedback, i.e. positive item-user pairs; $[z]_+ = \max(0, z)$ is the standard hinge loss; and $m > 0$ is the safety margin size. Notably, our loss function does not adopt the ranking loss weight compared to \cite{DBLP:conf/www/HsiehYCLBE17}.

\paragraph{Distortion Optimization.} The ultimate goal when embedding a space into another is to preserve distances while maintaining complex structures/relationships \cite{DBLP:conf/icml/SalaSGR18}. Thus, it becomes a challenge when embedding user-item pairs to hyperbolic space with the needs of preserving good structure quality for metric learning. To this end, we consider the two factors of good representations namely \textit{local} and \textit{global} factor. Locally, the children items must be spread out on the sphere around the parent user as described, with pull and push forces created by the gradients. Globally, the learned triplets should be separated reasonably from each other. While pull-and-push optimization satisfies the local requirement, we define the distortion optimization function to meet the global requirement as:
\begin{multline}
\mathcal{L}_{D} = \sum_{(i, j) \in \mathbb{S}} \Big[\frac{|d_\mathbb{D}(f(i), f(j)) - d_\mathbb{E}(i, j)|}{d_\mathbb{E}(i, j)}\Big]_+ \\ + \sum_{(i, k) \notin \mathbb{S}} \Big[\frac{|d_\mathbb{D}(f(i), f(k)) - d_\mathbb{E}(i, k)|}{d_\mathbb{E}(i, k)}\Big]_+,
\end{multline}
where $|\cdot|$ defines the absolute value; and $f(\cdot)$ is a mapping function $f: \mathbb{E} \rightarrow \mathbb{D}$ from Euclidean space $\mathbb{E}$ to hyperbolic space $\mathbb{D}$. In this paper, we take $f(\cdot)$ as an identity function.

We aim to preserve the distances by minimizing $\mathcal{L}_{D}$ for the global factor. Ideally, the lower the distortion, the better the preservation.  

\paragraph{Multi-Task Learning.} We then integrate the pull-and-push part (i.e., $\mathcal{L}_P$) and the distortion part (i.e., $\mathcal{L}_D$) into an end-to-end fashion through a multi-task learning framework. The objective function is defined as: 
\begin{equation}
\min_\Theta \mathcal{L} = \mathcal{L}_P + \gamma \mathcal{L}_D,
\end{equation}
where $\Theta$ is the total parameter space, including all embeddings and variables of the networks; and $\gamma$ is the multi-task learning weight. 

There is an unavoidable trade-off between the precision (learned from the pull-push mechanism) and the distortion as similar to \cite{DBLP:conf/icml/SalaSGR18}. Thus, jointly training $\mathcal{L}_P$ and $\mathcal{L}_D$ can help to boost the model performance while providing good representations. Indeed, we examine the performance of HyperML with and without the distortion by varying different multi-task learning weight $\gamma$ in our experiment in Section \ref{sec:experiments}.

\paragraph{Gradient Conversion.} The parameters of our model are learned by projected Riemannian stochastic gradient descent (RSGD) \cite{DBLP:conf/nips/NickelK17} with the form:
\begin{equation}
\boldsymbol{\theta}_{t+1} = \mathfrak{R}_{\theta_t}(-\eta_t\nabla_R \mathcal{L}(\boldsymbol\theta_t)),
\end{equation}
where $\mathfrak{R}_{\theta_t}$ denotes a retraction onto $\mathbb{D}$ at $\boldsymbol\theta$ and $\eta_t$ is the learning rate at time $t$.

The Riemannian gradient $\nabla_R$ is then calculated from the Euclidean gradient by rescaling $\nabla_E$ with the inverse of the Poincar\'e ball metric tensor as $\nabla_R = \frac{(1-\|\boldsymbol\theta_t\|^2)^2}{4} \nabla_E$, in which this scaling factor depends on the Euclidean distance of the point at time $t$ from the origin \cite{DBLP:conf/nips/NickelK17,DBLP:conf/wsdm/TayTH18}. Notably, one could also exploit full RSGD for optimization to perform the updates instead of using first-order approximation to the exponential map \cite{DBLP:journals/tac/Bonnabel13, DBLP:conf/icml/GaneaBH18,Wilson2018,DBLP:conf/iclr/BecigneulG19}.


\section{Experiments}
\label{sec:experiments}

\subsection{Experimental Setup}
\paragraph{Datasets.}

\begin{table}[t]
\centering
\resizebox{\linewidth}{!}{%
\begin{tabular}{c|cccc}
\hline \hline
\multicolumn{1}{c|}{\textbf{Dataset}} & \multicolumn{1}{c}{\textbf{Interactions}} & \multicolumn{1}{c}{\textbf{\# Users}} & \multicolumn{1}{c}{\textbf{\# Items}} & \multicolumn{1}{c}{\textbf{\% Density}} \\ \hline
Movie20M & 16M & 53K & 27K & 1.15 \\
Movie1M & 1M & 6K & 4K & 4.52 \\
Goodbooks & 6M & 53K & 10K & 1.14 \\
Yelp & 1M & 22K & 18K & 0.26 \\
Meetup & 248K & 47K & 17K & 0.03 \\
Clothing & 358K & 39K & 23K & 0.04 \\
Sports \& Outdoors & 368K & 36K & 18K & 0.06 \\
Cell Phones & 250K & 28K & 10K & 0.09 \\
Toys \& Games & 206K & 19K & 12K & 0.09 \\
Automotive & 26K & 3K & 2K & 0.49 \\ \hline \hline
\end{tabular}}
\caption{Statistics of all datasets used in our experimental evaluation}
\label{statistic}
\vspace{-7ex}
\end{table}

\begin{figure*}[t]
\begin{center}
\begin{minipage}[t]{3.9cm}
\includegraphics[width=3.9cm]{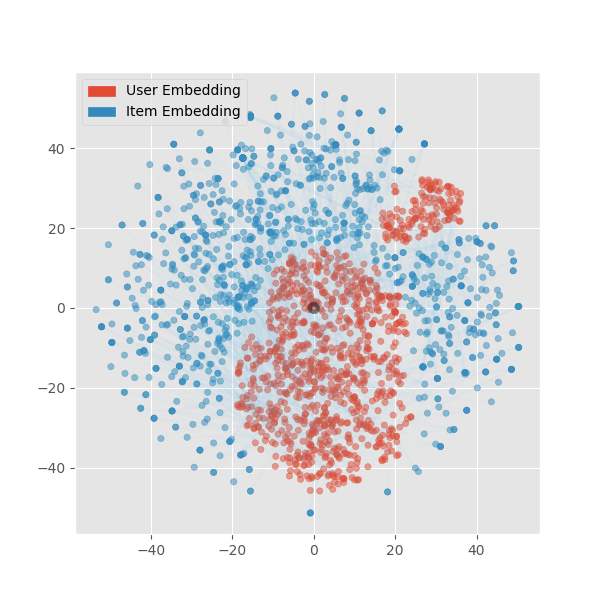}
\centering{MovieLens20M}
\end{minipage}
\begin{minipage}[t]{3.9cm}
\includegraphics[width=3.9cm]{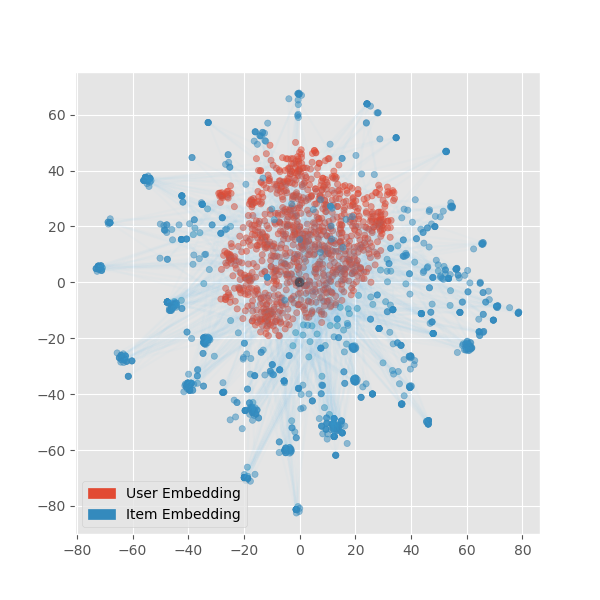}
\centering{MovieLens1M}
\end{minipage}
\begin{minipage}[t]{3.9cm}
\includegraphics[width=3.9cm]{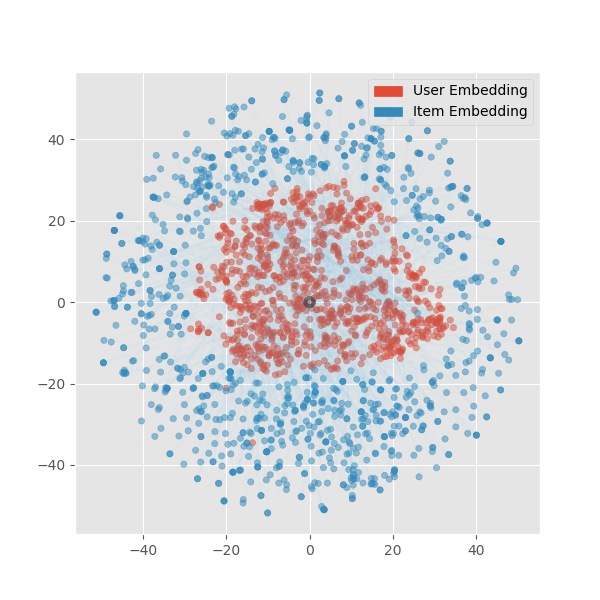}
\centering{Goodbooks}
\end{minipage}
\begin{minipage}[t]{3.9cm}
\includegraphics[width=3.9cm]{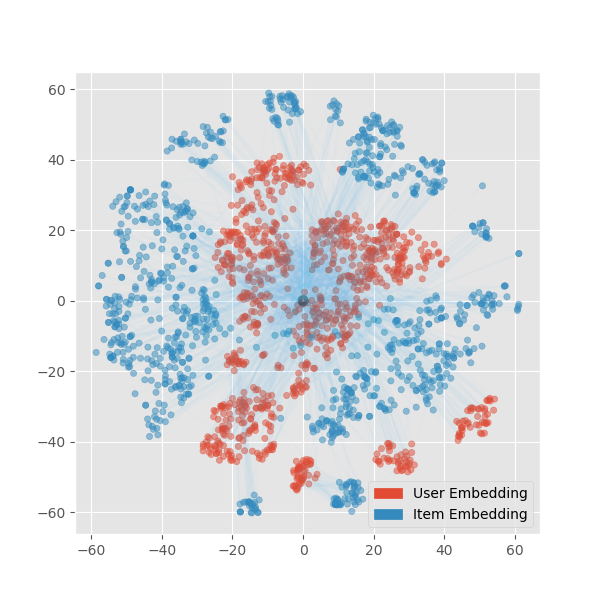}
\centering{Yelp}
\end{minipage}
\begin{minipage}[t]{3.9cm}
\includegraphics[width=3.9cm]{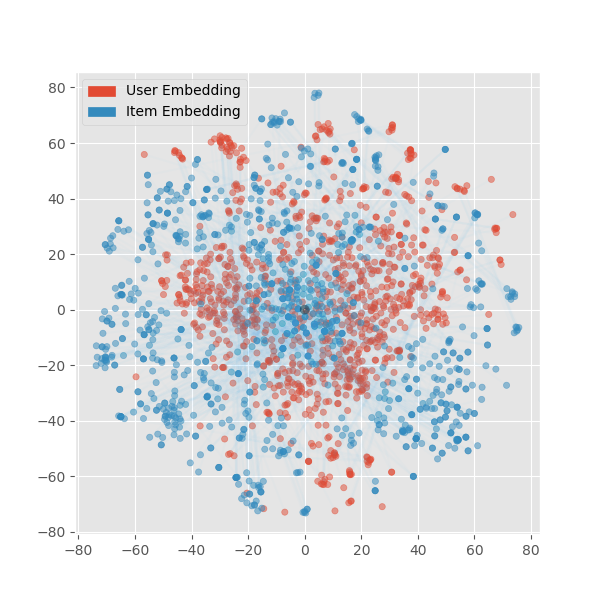}
\centering{Meetup}
\end{minipage}
\begin{minipage}[t]{3.9cm}
\includegraphics[width=3.9cm]{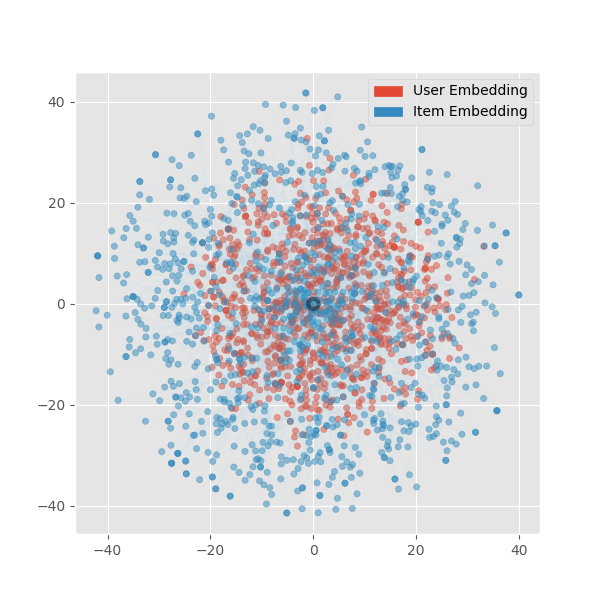}
\centering{Clothing, Shoes, and Jewelry}
\end{minipage}
\begin{minipage}[t]{3.9cm}
\includegraphics[width=3.9cm]{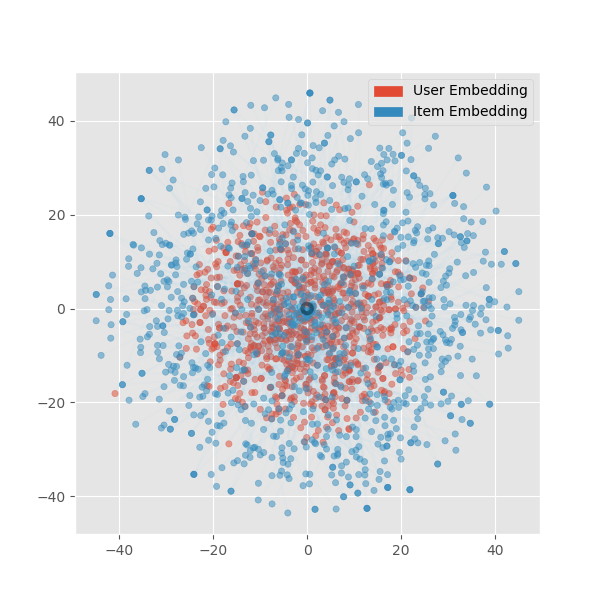}
\centering{Sports and Outdoors}
\end{minipage}
\begin{minipage}[t]{3.9cm}
\includegraphics[width=3.9cm]{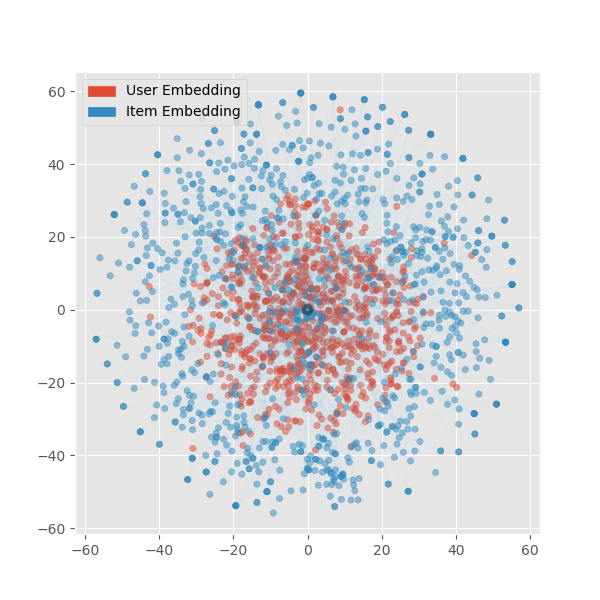}
\centering{Cell phones and Accessories}
\end{minipage}
\begin{minipage}[t]{3.9cm}
\includegraphics[width=3.9cm]{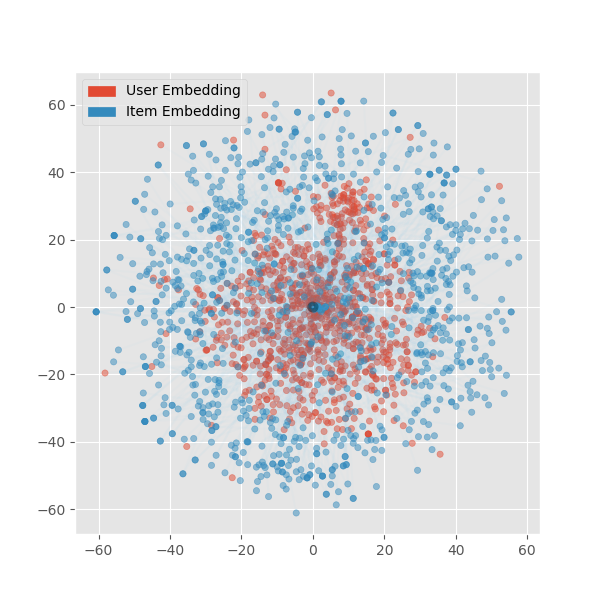}
\centering{Toys and Games}
\end{minipage}
\begin{minipage}[t]{3.9cm}
\includegraphics[width=3.9cm]{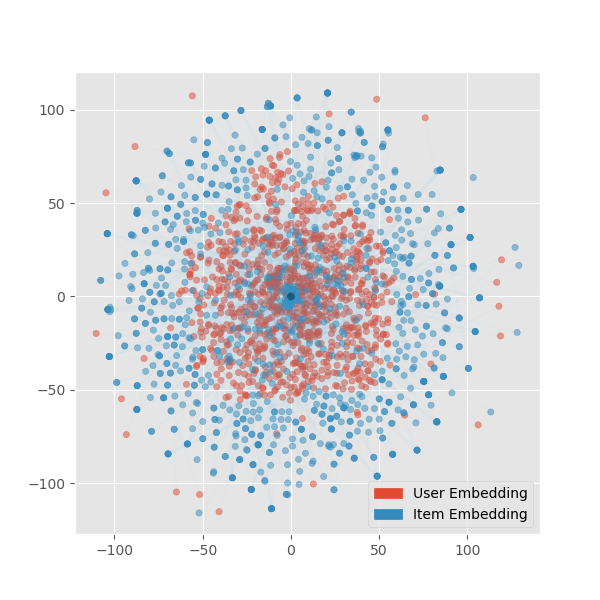}
\centering{Automotive}
\end{minipage}
\caption{Two-dimensional hyperbolic embedding of ten benchmark datasets in the Poincar\'e disk using t-SNE. The images illustrate the embedding of user and item pairs after the convergence (\textmd{\textit{Best viewed in color}}).}
\label{fig:hyperml}
\end{center}
\vspace{-4ex}
\end{figure*}

To evaluate our experiments, we use a wide spectrum of datasets with diverse domains and densities. The statistics of the datasets are reported in Table \ref{statistic}.

\begin{itemize}
    \item \textbf{MovieLens}: A widely adopted benchmark dataset in the application domain of recommending movies to users provided by GroupLens research\footnote{\url{https://grouplens.org/datasets/movielens/}}. We use two configurations, namely MovieLens20M and MovieLens1M. Similar to \cite{Tay:2018:LRM:3178876.3186154}, the MovieLens20M datasets are filtered with 100-core setting.
    \item \textbf{Goodbooks}: A large book recommendation dataset contains six million ratings for ten thousand most popular (with most ratings) books.\footnote{\url{https://github.com/zygmuntz/goodbooks-10k}}
    \item \textbf{Yelp}: A crowd-sourced platform for local businesses such as restaurants, bars, etc. We
    use the dataset from the 2018 edition of the Yelp Dataset Challenge.\footnote{\url{https://www.yelp.com/dataset/challenge}}
    \item \textbf{Meetup}:  An event-based social network dataset. We use the dataset includes event-user pairs from NYC that was provided by \cite{Pham:2016:GRM:3024719.3024759}.
    \item \textbf{Amazon Reviews}: The amazon review datasets that was introduced in \cite{DBLP:conf/www/HeM16}. The subsets\footnote{Datasets are obtained from \url{http://jmcauley.ucsd.edu/data/amazon/} using the 5-core setting with the domain names truncated in the interest of space.} are selected based on promoting diversity based on dataset size and domain.
\end{itemize}

\begin{table*}[t]
\centering
\resizebox{\linewidth}{!}{%
\begin{tabular}{l|cccccccccc}
\hline \hline
\multicolumn{1}{c|}{} & \multicolumn{2}{c}{MovieLens20M} & \multicolumn{2}{c}{MovieLens1M} & \multicolumn{2}{c}{Goodbooks} & \multicolumn{2}{c}{Yelp} & \multicolumn{2}{c}{Meetup} \\
 & \multicolumn{1}{c}{nDCG@10} & \multicolumn{1}{c}{HR@10} & \multicolumn{1}{c}{nDCG@10} & \multicolumn{1}{c}{HR@10} & \multicolumn{1}{c}{nDCG@10} & \multicolumn{1}{c}{HR@10} & \multicolumn{1}{c}{nDCG@10} & \multicolumn{1}{c}{HR@10} & \multicolumn{1}{c}{nDCG@10} & \multicolumn{1}{c}{HR@10} \\ \hline 
\begin{tabular}[c]{@{}l@{}}MF-BPR\end{tabular} & 63.462 & 82.206 & 55.173 & 74.057 & 49.559 & 71.033 & \underline{56.443} & \underline{77.926} & 48.359 & \underline{62.468} \\
\begin{tabular}[c]{@{}l@{}}MLP\end{tabular} & 62.500 & 84.380 & 54.851 & 73.812 & 48.597 & 70.226 & 52.777 & 75.784 & 43.310 & 55.616 \\
\begin{tabular}[c]{@{}l@{}}NCF\end{tabular} & 59.485 & 81.859 & 55.503 & 74.127 & \underline{50.823} & 72.014 & 53.078 & 72.757 & \underline{52.334} & 62.210 \\
\begin{tabular}[c]{@{}l@{}}CML \end{tabular} & 62.664 & \underline{85.571} & \underline{55.737} & \underline{74.528} & 49.010 & \underline{72.556} & 54.996 & 77.122 & 51.453 & 60.589 \\
\begin{tabular}[c]{@{}l@{}}LRML \end{tabular} & \underline{63.775} & 81.327 & 54.057 & 73.358 & 50.392 & 71.424 & 54.719 & 76.764 & 50.208 & 61.347 \\ 
\begin{tabular}[c]{@{}l@{}}HyperML\end{tabular} & \textbf{64.042} & \textbf{87.363} & \textbf{56.197} & \textbf{75.629} & \textbf{51.088} & \textbf{74.152} & \textbf{59.543} & \textbf{81.392} & \textbf{54.633} & \textbf{67.304} \\ \hline 
\begin{tabular}[c]{@{}l@{}}Improvement\end{tabular} & +0.42\% & +2.09\% & +0.83\% & +1.48\% & +0.52\% & +2.20\% & +5.49\% & +4.45\% & +4.39\% & +7.74\% \\ \hline \hline
\multicolumn{1}{c|}{} & \multicolumn{2}{c}{Clothing} & \multicolumn{2}{c}{Sports} & \multicolumn{2}{c}{Cell phones} & \multicolumn{2}{c}{Games} & \multicolumn{2}{c}{Automotive} \\
 & \multicolumn{1}{c}{nDCG@10} & \multicolumn{1}{c}{HR@10} & \multicolumn{1}{c}{nDCG@10} & \multicolumn{1}{c}{HR@10} & \multicolumn{1}{c}{nDCG@10} & \multicolumn{1}{c}{HR@10} & \multicolumn{1}{c}{nDCG@10} & \multicolumn{1}{c}{HR@10} & \multicolumn{1}{c}{nDCG@10} & \multicolumn{1}{c}{HR@10} \\ \hline
\begin{tabular}[c]{@{}l@{}}MF-BPR\end{tabular} & 13.189 & 20.509 & \underline{26.130} & \underline{38.553} & \underline{26.483} & 37.434 & \underline{22.156} & \underline{34.877} & \underline{20.433} & \underline{31.707} \\
\begin{tabular}[c]{@{}l@{}}MLP\end{tabular} & 13.947 & 22.726 & 24.431 & 37.015 & 25.732 & \underline{37.677} & 21.074 & 32.251 & 16.789 & 27.685 \\
\begin{tabular}[c]{@{}l@{}}NCF\end{tabular} & \underline{16.809} & \underline{26.470} & 20.268 & 30.232 & 22.496 & 32.697 & 20.959 & 30.871 & 17.340 & 28.441 \\
\begin{tabular}[c]{@{}l@{}}CML \end{tabular} & 16.623 & 26.371 & 19.211 & 30.197 & 19.320 & 29.746 & 21.579 & 32.524 & 17.556 & 27.985 \\
\begin{tabular}[c]{@{}l@{}}LRML \end{tabular} & 16.643 & 26.421 & 22.938 & 33.667 & 20.177 & 30.999 & 20.747 & 31.695 & 16.492 & 26.124 \\ 
\begin{tabular}[c]{@{}l@{}}HyperML\end{tabular} & \textbf{17.150} & \textbf{27.899} & \textbf{34.576} & \textbf{48.262} & \textbf{29.325} & \textbf{42.921} & \textbf{23.164} & \textbf{35.995} & \textbf{24.736} & \textbf{37.324} \\ \hline 
\begin{tabular}[c]{@{}l@{}}Improvement\end{tabular} & +2.03\% & +5.40\% & +32.32\% & +25.18\% & +10.73\% & +13.92\% & +4.55\% & +3.21\% & +21.06\% & +17.72\% \\ \hline \hline
\end{tabular}}
\caption{Experimental results (nDCG@10 and HR@10) on ten public benchmark datasets. Best result is in bold face and second best is underlined. Our proposed HyperML achieves very competitive results, outperforming strong recent advanced metric learning baselines such as CML and LRML.}
\label{exp_results}
\vspace{-6ex}
\end{table*}

\paragraph{Evaluation Protocol and Metrics.}
We experiment on the one-class collaborative filtering setup. We adopt nDCG@10 (normalized discounted cumulative gain) and HR@10 (Hit Ratio) evaluation metrics, which are well-established ranking metrics for recommendation task. Following \cite{DBLP:conf/www/HeLZNHC17,Tay:2018:LRM:3178876.3186154}, we randomly select 100 negative samples which the user have not interacted with and rank the ground truth amongst these negative samples. For all datasets, the last item the user has interacted with is withheld as the test set while the penultimate serves as the validation set. During training, we report the test scores of the model based on the best validation scores. All models are evaluated on the validation set at every 50 epochs.

\vspace{-1.75ex}
\paragraph{Compared Baselines}
In our experiments, we compare with five well-established and competitive baselines which in turn employ different matching functions: inner product (MF-BPR), neural networks (MLP, NCF) and metric learning (CML, LRML).

\vspace{-2ex}
\begin{itemize}
\item \textbf{Matrix Factorization with Bayesian Personalized Ranking (MF-BPR)} \cite{DBLP:conf/uai/RendleFGS09} is the standard and strong collaborative filtering (CF) baseline for recommender systems. It models the user-item representation using the inner product and explores the triplet objective to rank items.
\item \textbf{Multi-layered Perceptron (MLP)} is a feedforward neural network that applies multiple layers of nonlinearities to capture the relationship between users and items. We select the best number of MLP layers from $\{3, 4, 5\}$. 
\item \textbf{Neural Collaborative Filtering (NCF)} \cite{DBLP:conf/www/HeLZNHC17} is a neural network based method for collaborative filtering which models nonlinear user-item interaction. The key idea of NCF is to fuse the last hidden representation of MF and MLP into a joint model. Following \cite{DBLP:conf/www/HeLZNHC17}, we use a three layered MLP with a pyramid structure.
\item \textbf{Collaborative Metric Learning (CML)} \cite{DBLP:conf/www/HsiehYCLBE17} is a strong metric learning baseline that learns user-item similarity using the Euclidean distance. CML can be considered a key ablative baseline in our experiments, signifying the difference between Hyperbolic and Euclidean metric spaces.
\item \textbf{Latent Relational Metric Learning (LRML)} \cite{Tay:2018:LRM:3178876.3186154} is also a strong metric learning baseline that learns adaptive relation vectors
between user and item interactions to find a single optimal translation vector between each user-item pair.
\end{itemize}

\vspace{-3ex}
\paragraph{Implementation Details}
We implement all models in Tensorflow. All models are trained with the Adam \cite{DBLP:journals/corr/KingmaB14} or AdaGrad \cite{DBLP:journals/jmlr/DuchiHS11} optimizer with learning rates from $\{0.01, 0.001, 0.0001, 0.00001\}$. The embedding size $d$ of all models is tuned amongst $\{32, 64, 128\}$ and the batch size $B$ is tuned amongst $\{128, 256, 512\}$. The multi-task learning weight $\gamma$ is empirically chosen from  $\{0, 0.1, 0.25, 0.5, 0.75, \\1.0\}$. For models that optimize the hinge loss, the margin $\lambda$ is selected from $\{0.1, 0.2, 0.5\}$. For NCF, we use a pre-trained model as reported in \cite{DBLP:conf/www/HeLZNHC17} to achieve its best performance. All the embeddings and parameters are randomly initialized using the random uniform initializer $\mathcal{U}(-\alpha, \alpha)$. For non-metric learning baselines, we set $\alpha = 0.01$. For metric learning models, we empirically set $\alpha = (\frac{3\beta^2}{2d})^{\frac{1}{3}}$, in which we choose $\beta = 0.01$. The reason is that we would like all the embeddings of the metric learning models to be initialized arbitrarily close to the origin of the balls\footnote{The balls are referred as hyperbolic ball for HyperML model and Euclidean ball for CML and LRML model.} for a fair comparison. For most datasets and baselines, we empirically set the embedding size of $64$ and the batch size is $512$. We also empirically set the dropout rate $\rho=0.5$ to prevent overfitting. For each dataset, the optimal parameters are established by repeating each experiment for $N$ runs and assessing the average results. We have used $N = 5$ for our experiment.

\vspace{-1ex}
\subsection{Experimental Results}
This section presents our experimental results on all datasets. For all obtained results, the best result is in boldface whereas the second best is underlined. As reported in Table \ref{exp_results}, our proposed model consistently outperforms all the baselines on both HR@10 and nDCG@10 metrics across all benchmark datasets.

Pertaining to the baselines, we observe that there is no obvious winner among the baseline solutions. In addition, we also observe that the performance of MF-BPR and CML is extremely competitive, i.e. both MF-BPR and CML consistently achieve good results across the datasets. Notably, the performance of MF-BPR is much better than CML on datasets with less number of interactions. For datasets with larger size (i.e., MovieLens20M, MovieLens1M and Goodbooks), the performance of metric learning models perform better in which the gain of CML and LRML on the non-metric learning baselines across large datasets is approximately +0.39\% and +0.91\% respectively in terms of nDCG. One possible reason is that for small datasets with low interactions (e.g., Automotive with 26K interactions of $0.49\%$ density), a simple model such as MF-BPR should be considered as a priority choice. In addition, the performance of a careful pre-trained NCF also often achieves competitive results with large datasets but not small ones in most cases. The explanation is because using the dual embedding spaces (since NCF combines MLP and MF) could possibly lead to overfitting if the dataset is not large enough \cite{Tay:2018:LRM:3178876.3186154}.

Remarkably, our proposed model HyperML demonstrates highly competitive results and consistently outperforms the best baseline method. The percentage improvements in terms of nDCG on ten datasets (in the same order as reported in Table \ref{exp_results}) are +0.42\%, +0.83\%, +0.52\%, +5.49\%, +4.39\%, +2.03\%, +32.32\%, +10.71\%, +4.55\% and +21.06\% respectively. We also observe similar high performance gains on the hit ratio (HR@10). Note that the hyperbolic spaces expand faster, i.e. exponentially, than Euclidean spaces, in which the forces are generated by the rescaled gradients, pulling and pushing the points with more reasonable distances compared to Euclidean. Therefore, it enables us to achieve very competitive results of our proposed HyperML in the hyperbolic space over other strong Euclidean baselines. Our experimental evidence shows the remarkable recommendation results of our proposed HyperML model on the variety of datasets and the advantage of hyperbolic space over Euclidean space in boosting the performance in metric learning framework.

\vspace{-1ex}
\subsection{Model Convergence Analysis}
This section investigates the model convergence analysis of our proposed model to understand the behavior of the embeddings in hyperbolic space.

\paragraph{Hyperbolic Convergence}
Figure \ref{fig:hyperml} visualizes the two-dimensional hyperbolic embedding using t-SNE on the test set of ten benchmark datasets after the convergence. We observe that item embeddings form a sphere over the user embeddings. Moreover, since we conduct the analysis on the \textbf{test} set, the visualization of the user/item embeddings in Figure \ref{fig:hyperml} demonstrates the ability of HyperML to self-organize and automatically spread out the item embeddings on the sphere around user embeddings, as mentioned in \cite{Tay:2018:LRM:3178876.3186154,DBLP:conf/icml/SalaSGR18,DBLP:conf/iclr/TifreaBG19}. Moreover, the clustering characteristic of observing the user-user and item-item relationships discussed in Section \ref{sec:hyperbolic_metric_learning} is also captured in Figure \ref{fig:hyperml}. It could be seen especially clearly for the MovieLens and Yelp dataset.

\begin{figure}[t]
\begin{center}
\begin{minipage}[t]{2.7cm}
\includegraphics[width=2.7cm]{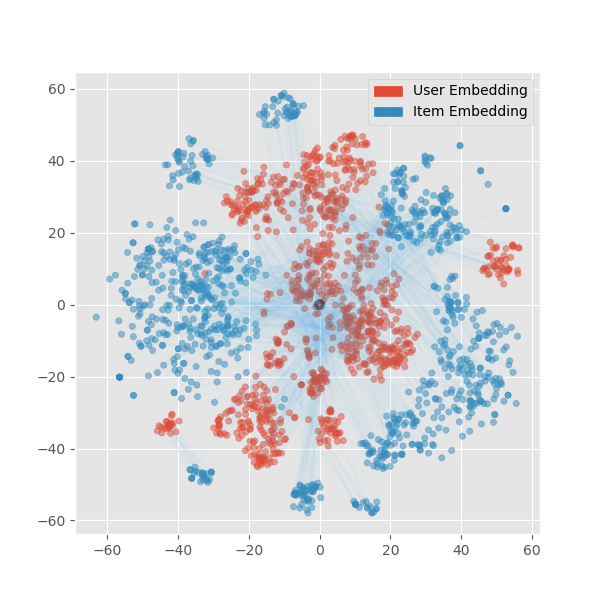}
\centering{LRML}
\end{minipage}
\begin{minipage}[t]{2.7cm}
\includegraphics[width=2.7cm]{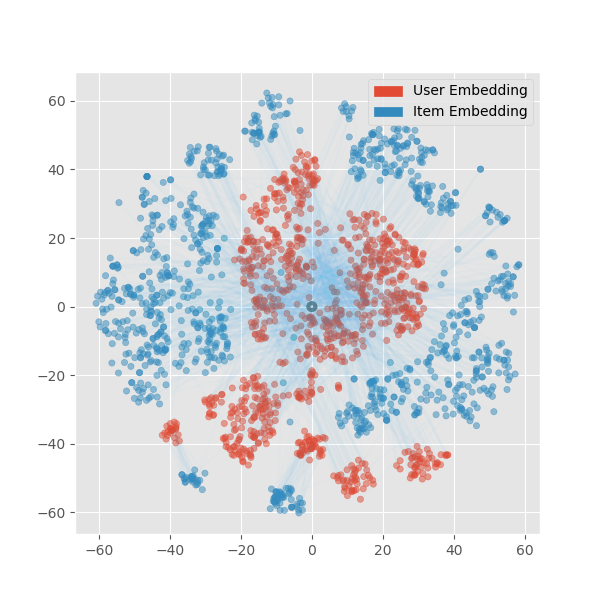}
\centering{CML}
\end{minipage}
\begin{minipage}[t]{2.7cm}
\includegraphics[width=2.7cm]{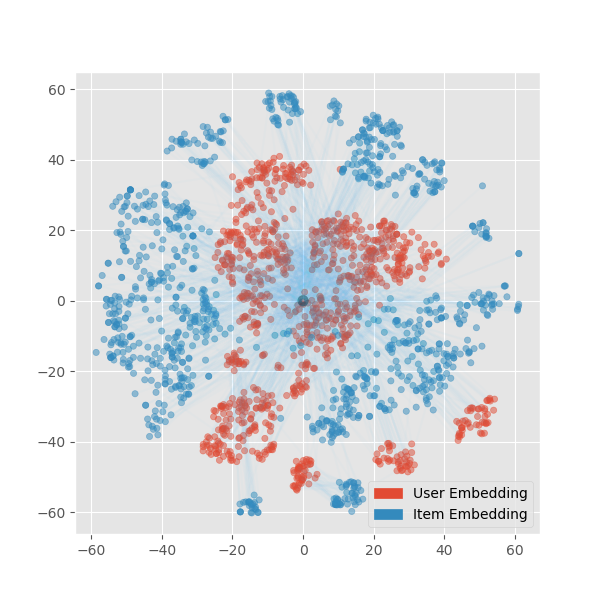}
\centering{HyperML}
\end{minipage}
\end{center}
\centering{(a). Embeddings comparison on Yelp dataset.}
\vspace{-3.95ex}
\end{figure}

\begin{figure}[t]
\begin{center}
\begin{minipage}[t]{2.7cm}
\includegraphics[width=2.7cm]{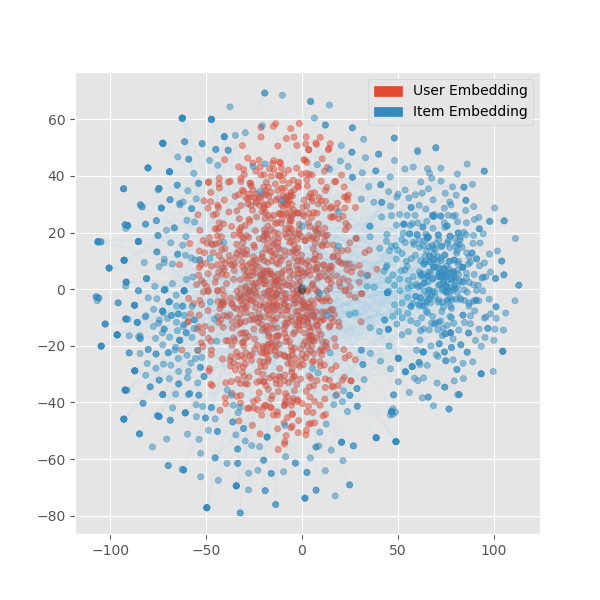}
\centering{LRML}
\end{minipage}
\begin{minipage}[t]{2.7cm}
\includegraphics[width=2.7cm]{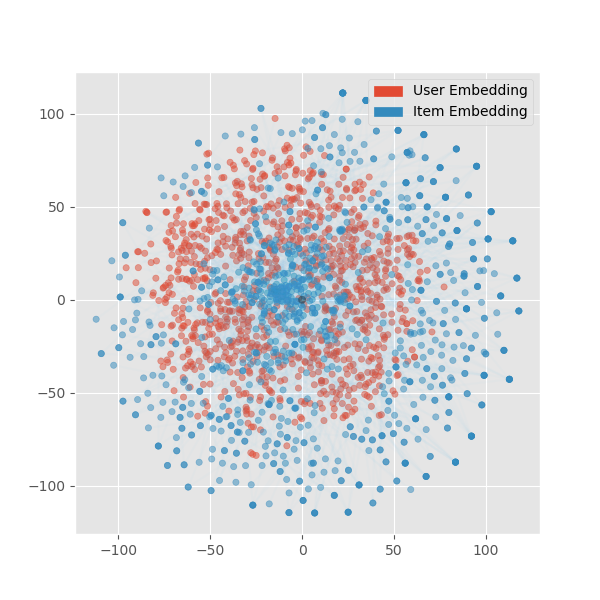}
\centering{CML}
\end{minipage}
\begin{minipage}[t]{2.7cm}
\includegraphics[width=2.7cm]{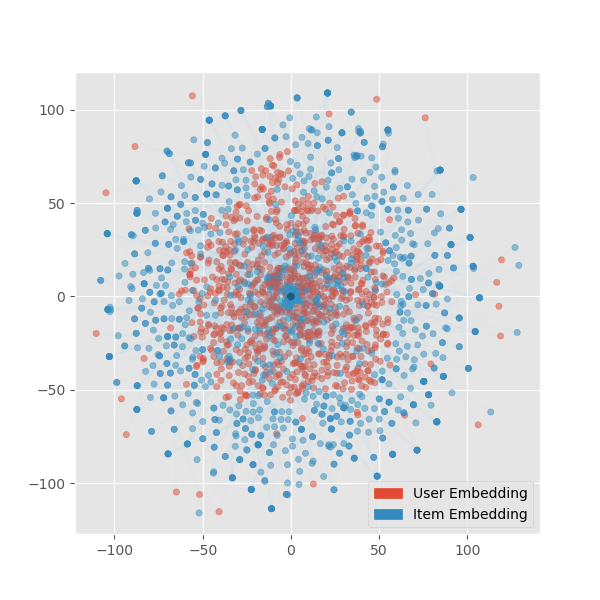}
\centering{HyperML}
\end{minipage}
\centering{(b). Embeddings comparison on Automotive dataset.}
\caption{Comparison between two-dimensional Poincar\'e embedding and Euclidean embedding on Yelp and Automotive dataset. The images illustrate the embeddings of LRML, CML and HyperML after convergence (\textmd{\textit{Best viewed in color}}).}
\label{fig:converge_comparison}
\end{center}
\vspace{-4.5ex}
\end{figure}


\paragraph{Convergence Comparison} 
Figure \ref{fig:converge_comparison} illustrates the comparison between two-dimensional Poincar\'e embedding (HyperML) and Euclidean embedding (CML, LRML) on the Yelp and Amazon dataset. For the Euclidean embedding, we clip the norm (i.e., the norm of the embeddings is constrained to 1) and initialize all the embeddings very close to the origin, for an analogous comparison. 

At first glance, we notice the difference between the three types of embedding by observing the distribution of user and item embeddings in the spaces after the convergence. While HyperML and CML have the item embeddings gradually assemble to a sphere structure surround user embeddings, the item embeddings of LRML have the opposite movement. The reason is because the motivation behind both CML and HyperML is to create the learned metric through the pull-push mechanism, whereas the motivation of LRML is to additionally learn the translation vector, which establishes the main cause of different visualizations. 

It is worth mentioning that since we initialize all the embeddings very close to the origin, we observe the difference between hyperbolic and Euclidean space that leads to the difference in the convergence of HyperML and CML. While both models form a sphere shape over the user embeddings equally, we observe that the user embeddings of HyperML tend to be located closer to the origin than CML while we get similar spread out observation of items. The explanation is that even with similar forces created by the gradients in the same direction, the different expansion property of the two spaces produces the distances between the triplets more separable, which leads to the boosting performance of the proposed model. 

\vspace{-2ex}
\subsection{Comparison of Hyperbolic Variants}
In this section, we study the variants of our proposed model: HyperML and HyperTS (applied optimization after mapping the user and item embeddings to the tangent space at $\textbf{0}$ using the $\log_\textbf{0}$ map). Notably, HyperTS is viable because the tangent space at the origin of the Poincar\'e ball resembles Euclidean space. Table \ref{comparison_variants} represents the performance of the variants on the datasets in terms of nDCG@10 and HR@10. In general, we observe both HyperML and HyperTS achieve highly competitive results, boosting the performance over Euclidean metric learning models across Meetup, Clothing, Sports, and Cell phones datasets.

\begin{table}[t]
\centering
\begin{tabular}{l|cccccc}
\hline \hline
\multicolumn{1}{c|}{} & \multicolumn{2}{c}{Meetup} & \multicolumn{2}{c}{Clothing} \\
 & \multicolumn{1}{c}{nDCG@10} & \multicolumn{1}{c}{HR@10} & \multicolumn{1}{c}{nDCG@10} & \multicolumn{1}{c}{HR@10} \\ \hline
\begin{tabular}[c]{@{}l@{}}HyperML\end{tabular} & 54.633 & 67.304 & 17.150 & 27.899 \\
\begin{tabular}[c]{@{}l@{}}HyperTS\end{tabular} & 54.612 & 67.277 & 17.190 & 27.959 \\ \hline \hline
\multicolumn{1}{c|}{} & \multicolumn{2}{c}{Sports} & \multicolumn{2}{c}{Cell phones} \\
 & \multicolumn{1}{c}{nDCG@10} & \multicolumn{1}{c}{HR@10} & \multicolumn{1}{c}{nDCG@10} & \multicolumn{1}{c}{HR@10} \\ \hline
\begin{tabular}[c]{@{}l@{}}HyperML\end{tabular} & 34.576 & 48.262 & 29.325 & 42.921 \\
\begin{tabular}[c]{@{}l@{}}HyperTS\end{tabular} & 31.896 & 45.272 & 29.933 & 43.532 \\ \hline \hline
\end{tabular}
\caption{Performance comparison between HyperML and HyperTS.}
\label{comparison_variants}
\vspace{-6ex}
\end{table}

\vspace{-1ex}
\subsection{Effect of Scaling Variable}
In this section, we study the effect of the variable $c$ on our proposed HyperML and the CML baseline model. Table \ref{effect_of_scaling} represents the performance of HyperML regarding the different value of the scaling variable $c$ comparing to CML in terms of nDCG@10. We observe that HyperML achieves best performance when $c=0.5$ on Goodbooks dataset, but $c=2.0$ on Games dataset. For other values of $c$, we notice the oscillated performance of HyperML. 

As introduced, for $c>0$, our ball shrinks to the radius of $\frac{1}{\sqrt{c}}$. Without loss of generality, the case of $c>0$ can be reduced to $c=1$ (the usual unit ball). However, we observe that different scaling variable $c$ would effect the performance differently in practice, which should be set carefully for each dataset. In fact, with $c$ carries values from 0.5 to 8.0, the percentage gain/loss of HyperML over CML varies from $+4.87\%/-17.39\%$ to $+14.17\%/-12.71\%$ on Goodbooks and Games dataset, respectively.

\begin{table}[t]
\centering
\resizebox{\linewidth}{!}{%
\begin{tabular}{c|cccccc}
\hline \hline
\multicolumn{1}{c|}{Scaling} & \multicolumn{3}{c}{Goodbooks} & \multicolumn{3}{c}{Games} \\
 \multicolumn{1}{c|}{Variable $c$}
 & \multicolumn{1}{c}{HyperML} & \multicolumn{1}{c}{CML} & \multicolumn{1}{c}{$\Delta\textit{(\%)}$} & \multicolumn{1}{c}{HyperML} & \multicolumn{1}{c}{CML} & \multicolumn{1}{c}{$\Delta\textit{(\%)}$} \\ \hline 
$c=0.5$ & 51.396 & 49.010 & $+4.87\%$ & 37.134 & 32.524 & $+14.17\%$ \\
$c=1.0$ & 51.088 & 49.010 & $+4.24\%$ & 35.995 & 32.524 & $+10.67\%$ \\
$c=2.0$ & 49.631 & 49.010 & $+1.27\%$ & 38.578 & 32.524 & $+18.61\%$ \\
$c=4.0$ & 46.017 & 49.010 & $-6.11\%$ & 35.466 & 32.524 & $+9.05\%$ \\
$c=8.0$ & 40.488 & 49.010 & $-17.39\%$ & 28.390 & 32.524 & $-12.71\%$ \\ \hline \hline
\end{tabular}}
\caption{Effects of the scaling variable $c$ on Goodbooks and Games datasets in terms of nDCG@10.}
\label{effect_of_scaling}
\vspace{-6ex}
\end{table}

\subsection{Accuracy Trade-off with Different Multi-task Learning Weight}
\begin{figure}[t]
\begin{center}
\begin{minipage}[t]{4.2cm}
\includegraphics[width=4.2cm]{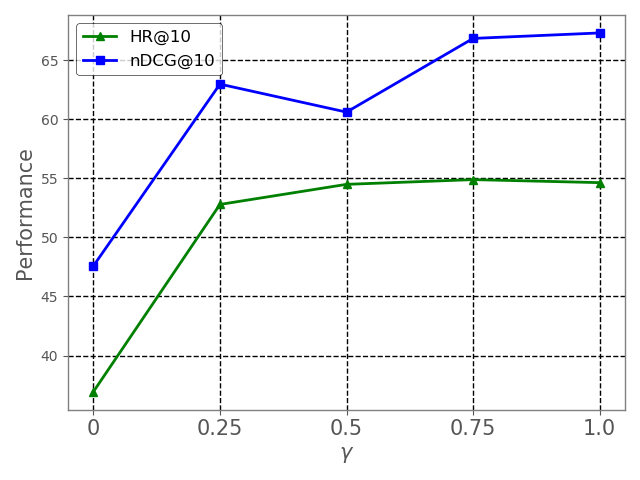}
\centering{(a) Meetup}
\end{minipage}
\begin{minipage}[t]{4.2cm}
\includegraphics[width=4.2cm]{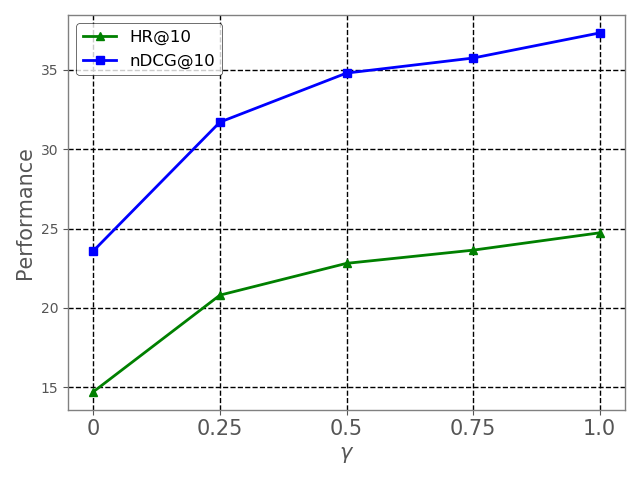}
\centering{(b) Automotive}
\end{minipage}
\caption{Performance on Different Multi-task Learning Weight $\gamma$.}
\label{fig:multi_task}
\end{center}
\vspace{-4ex}
\end{figure}

In this section, we study the effect of different multi-task learning weight $\gamma$ on our proposed HyperML model on Meetup and Automotive dataset. Figure \ref{fig:multi_task} represents the performance of HyperML when changing the value of the multi-task learning weight $\gamma$ in terms of HR@10 and nDCG@10. We observe the obvious boost of the performance when $\gamma$ increases from 0 to positive values on both two datasets. While for Meetup dataset, HyperML achieves best performance when $\gamma=0.75$, we observe the performance of HyperML achieves its best result on Automotive dataset when $\gamma=1.0$. For other values of $\gamma$, we also observe the oscillated performance of HyperML due to the trade-off. On a side note, when $\gamma=0$, i.e. removing the distortion, we notice the decreasing performance of HyperML compared to CML by -21.48\% and -15.48\% in terms of nDCG@10 on Meetup and Automotive dataset, respectively.

Thus, we conclude that the multi-task learning weight $\gamma$ as well as the distortion play important roles on boosting the performance, in which the weight $\gamma$ causes the trade-off between minimizing the distortion and the model's accuracy.

\begin{figure}[t]
\centering
\includegraphics[width=0.4\textwidth]{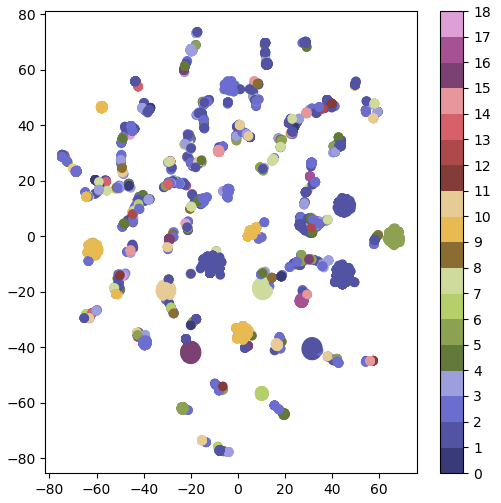}
\caption{2D t-SNE item embeddings visualization of hyperbolic metric in MovieLens1M dataset (\textmd{\textit{Best viewed in color}}).}
\label{fig:metric_visualization}
\vspace{-3ex}
\end{figure}

\subsection{Metric Learning Visualization}
In this section, we study the clustering effect of HyperML. Figure \ref{fig:metric_visualization} represents the clustering effect in which the 18 colors represent 18 movie genres from the MovieLens1M dataset\footnote{The colors were assigned to the movie genres in the same order as reported in \url{http://files.grouplens.org/datasets/movielens/ml-1m-README.txt}}. From the figure, we empirically discover that despite being only trained on implicit interactions, explicit rating information is surprisingly being discovered in HyperML. Within the hyperbolic space, the metric learning shows cluster structures of items with same genres induced by users, providing insight and achieving similar effect as \cite{He:2016:VVB:3015812.3015834,DBLP:conf/www/HsiehYCLBE17}. The visualization supports our previous claim that the nearest neighborhood items tend to be liked by the same users with similar interests. Notably, the t-SNE visualization also illustrates the sphere structure embeddings as introduced.

\vspace{-1.5ex}
\section{Related Work}
Across the rich history of recommender systems research, a myriad of machine learning models have been proposed using matching functions to define similarity scores \cite{DBLP:conf/uai/RendleFGS09,DBLP:conf/icdm/Rendle10,mnih2008probabilistic,DBLP:conf/uai/RendleFGS09,he2016fast,koren2008factorization,DBLP:conf/www/HeLZNHC17,DBLP:conf/www/HsiehYCLBE17}. Traditionally, many works are mainly focused on factorizing the interaction matrix, combining the user-item embeddings using the inner product as a matching function \cite{mnih2008probabilistic,koren2009matrix,DBLP:conf/www/HeLZNHC17}. On the other hand, many approaches in personalized recommender system based on the distance/similarity metric between two points using Euclidean distance have shown their strong competency in improving the model accuracy in different domains \cite{DBLP:conf/nips/WeinbergerBS05,DBLP:conf/nips/WangDWK11,DBLP:conf/cvpr/ChopraHL05,DBLP:conf/nips/KedemTWSL12,DBLP:conf/nips/XingNJR02,DBLP:conf/www/TranLLK19,DBLP:conf/sigir/TranSL19}.

To this end, \cite{DBLP:conf/www/HsiehYCLBE17} argued that using inner product formulation lacks expressiveness due to its violation of the triangle inequality. As a result, the authors proposed Collaborative Metric Learning (CML), a strong recommendation baseline based on Euclidean distance. Notably, many recent works have moved into neural models \cite{DBLP:conf/www/HeLZNHC17,ijcai2018-510}, in which stacked nonlinear transformations have been used to approximate the interaction function. 

Our work is inspired by recent advances in hyperbolic representation learning \cite{DBLP:conf/nips/NickelK17,pmlr-v89-cho19a,DBLP:conf/icml/NickelK18,DBLP:conf/icml/GaneaBH18,DBLP:conf/icml/SalaSGR18,DBLP:conf/uai/DavidsonFCKT18,DBLP:conf/iclr/TifreaBG19,DBLP:conf/iclr/GuSGR19,DBLP:conf/icml/LawLSZ19}. For instance, \cite{DBLP:conf/wsdm/TayTH18} proposed training a question answering system in hyperbolic space. \cite{DBLP:conf/textgraphs/DhingraSNDD18} proposed learning word embeddings using a hyperbolic neural network. \cite{DBLP:conf/iclr/GulcehreDMRPHBB19} proposed a hyperbolic variation of self-attention and the transformer network, and applied it to tasks such as visual question answering and neural machine translation. \cite{DBLP:conf/nips/GaneaBH18} proposed recurrent neural networks in hyperbolic space, \cite{DBLP:journals/corr/ChamberlainCD17} proposed a method of embedding graphs in hyperbolic space. \cite{DBLP:journals/corr/abs-1902-08648} is the most similar work to ours that embeds bipartite user-item graphs in hyperbolic space, but it does not learn the embeddings with metric learning manner. While the advantages of hyperbolic space seem eminent in the wide variety of application domains, there is no work that investigates this embedding space within the context of metric learning in recommender systems. This constitutes the key novelty of our work. A detailed primer on hyperbolic space is given in the technical exposition of the paper.

\vspace{-2ex}
\section{Conclusion}
In this paper, we introduce a new effective and competent recommendation model called HyperML. To the best of our knowledge, HyperML is the first model to explore metric learning in hyperbolic space in recommender system. Additionally, we also introduce a distortion term, which is essential to control good representations in hyperbolic space. Through extensive experiments on ten datasets, we are able to demonstrate the effectiveness of HyperML over other baselines in Euclidean space, even state-of-the-art metric learning models such as CML or LRML. The promising results of HyperML may inspire other future works to explore hyperbolic space in solving recommendation problems.

\bibliographystyle{ACM-Reference-Format}
\bibliography{references}

\end{document}